# FORM FACTORS AND QCD IN SPACELIKE AND TIMELIKE REGIONS

## A. P. Bakulev

*Bogolyubov Laboratory of Theoretical Physics, JINR, Dubna, 141980 Russia*

## A. V. Radyushkin[†]

*Physics Department, Old Dominion University, Norfolk, VA 23529, USA*
*and*
*Theory Group, Jefferson Lab, Newport News, VA 23606, USA*

## N. G. Stefanis

*Institut für Theoretische Physik II, Ruhr-Universität Bochum, D-44780 Bochum, Germany*

We analyze the basic hard exclusive processes: $\pi\gamma^*\gamma$-transition, pion and nucleon electromagnetic form factors, and discuss the analytic continuation of QCD formulas from the spacelike $q^2 < 0$ to the timelike region $q^2 > 0$ of the relevant momentum transfers. We describe the construction of the timelike version of the coupling constant $\alpha_s$. We show that due to the analytic continuation of the collinear logarithms each eigenfunction of the evolution equation acquires a phase factor and investigate the resulting interference effects which are shown to be very small. We found no sources for the $K$-factor-type enhancements in the perturbative QCD contribution to the hadronic form factors. To study the soft part of the pion electromagnetic form factor, we use a QCD sum rule inspired model and show that there are non-canceling Sudakov double logarithms which result in a $K$-factor-type enhancement in the timelike region.

## I. INTRODUCTION

Within the factorization framework, perturbative QCD has been applied to various processes involving large momentum transfers, both in the spacelike $q^2 = -Q^2 < 0$ (for reviews, we refer to [1–6]) and the timelike $q^2 > 0$ regions (see, for example, [7–10]). Note that the running coupling constant $\alpha_s(\mu^2)$ is usually defined with reference to some Euclidean (spacelike) configuration of momenta of scale $\mu$. For large spacelike $q$, this produces no special complications. One simply uses the renormalization group to resum the logarithmic corrections $(\alpha_s(\mu^2)\ln(Q^2/\mu^2))^N$ that appear in higher orders of perturbation theory, arriving at an expansion in the effective coupling constant $\alpha_s(Q^2)$ which, in the 1-loop approximation, is given by [1]

$$\alpha_s(Q^2) = \frac{4\pi}{(11 - 2N_f/3)\ln(Q^2/\Lambda^2)} \, , \tag{1}$$

with $N_f$ being the number of active flavors and $\Lambda$ denoting $\Lambda^{\rm QCD}$. In general, the $\Lambda$-parameterization of $\alpha_s(Q^2)$ is a series expansion in $1/L$ (where $L = \ln(Q^2/\Lambda^2)$), and the definition of $\Lambda$ is fixed only if the $O(1/L^2)$-term is added to Eq.(1) [11]. Continuing the logarithms into the region of timelike $q$, one should deal with the $i\pi$ terms:

---

[†]Also Laboratory of Theoretical Physics, JINR, Dubna, 141980 Russia

$\ln(Q^2/\mu^2) \to \ln(Q^2/\mu^2) \pm i\pi$, which may produce large higher-order corrections. In the case of the $R$-ratio for $e^+e^- \to hadrons$ process, this problem was discussed in refs. [12–14]. It was shown there that, by using the $\Lambda$-parameterization for $\alpha_s(Q^2)$ in the spacelike region, it is possible to construct for $R(q^2)$ an expansion in the timelike region in which all the $(\pi^2/L^2)^N$-terms are resummed explicitly, and, what is most important, the transformation into the timelike region *reduces* the magnitude of each particular term of the $1/L$ expansion. Another well-studied example related to the analytic continuation into the timelike region is the cross section of the Drell-Yan (DY) process $AB \to \gamma^* X$. In this case, the $i\pi$ factors associated with the continuation of the Sudakov double logarithms $(\alpha_s \ln^2(Q^2/m^2))^N$ result in a $\pi^2$-enhanced correction which gives rise to the $K$-factor [15] increasing in turn the result of the perturbative QCD calculation by the factor of 3 to bring the DY cross section in agreement with experiment.

For elastic form factors, existing experimental data [16–18] show a considerable enhancement of the timelike form factors over their spacelike counterparts. In the present paper, we study possible sources of such an enhancement. To disentangle different aspects of the analytic continuation into the timelike region, we proceed step by step, beginning with the simplest cases and then going on to more complicated ones. We start with a discussion of the analytic continuation into the timelike region of the UV logarithms $\ln(Q^2/\mu_R^2)$ inducing the $Q^2$-dependence of the running coupling constant $\alpha_s(Q^2)$. We take the cleanest case of $R(e^+e^- \to hadrons)$, in which no other types of logarithms appear and review in Section II the continuation procedure for $R(e^+e^- \to hadrons)$ as given in refs. [13,14]. In section III, we consider another fundamental process: $\gamma^*\gamma \to \pi^0$. At the leading logarithm level, only the collinear logarithms $\ln(Q^2/\mu_F^2)$ are important while $\alpha_s$ can be treated as a constant. So, this is another "clean situation" which gives an opportunity to concentrate on the study of the analytic continuation of the collinear logarithms which induce the $Q^2$-dependence of the pion distribution amplitude $\varphi_\pi(x, Q^2)$. In Section IV, we briefly discuss the effects due to the analytic continuation of the Sudakov double logarithms. We consider first the cross section of the Drell-Yan process $AB \to \mu^+\mu^- X$. In this case, the double logs $\ln^2(Q^2/\mu^2)$ appear on a diagram by diagram basis but cancel after resumming over all diagrams of a given order. However, the $\pi^2$ terms generated by the analytic continuation survive and, as already mentioned, produce an enhancement due to the $K$-factor. We contrast this outcome with the case of the hard contribution to the pion electromagnetic form factor, in which the induced $\pi^2$ terms cancel together with the double logs. For this reason, the modification of the hard term of the pion form factor in the timelike region is only affected by the analytic continuation of the UV and collinear logarithms. These effects are discussed in Section V. In Section VI, we study the analytic continuation of the hard pQCD contribution to the nucleon form factor. Both in the pion and the nucleon case, we find that the effects due to the continuation into the timelike region are very small. Experimentally, however, the timelike nucleon form factor is essentially larger than its spacelike counterpart. This discrepancy may be regarded as an indication that the hard contribution does not dominate the form factors at accessible momentum transfers. An alternative scenario discussed in many papers [19–26] is that in the few GeV$^2$ region the form factors are dominated by the soft mechanism. In Section VII, we study the analytic continuation effects for the soft contribution to the pion electromagnetic form factor within the local quark-hadron duality model motivated by the QCD sum rule analysis of refs. [19–21,26]. We show that at the one loop level, there are explicit non-canceled double logarithms $\ln^2(Q^2/\mu^2)$ which produce the $\pi^2$ terms in the timelike region, giving rise to a $K$-factor-type enhancement.

## II. CONTINUATION OF $\alpha_S$ INTO THE TIMELIKE REGION: $R(e^+e^- \to hadrons, s)$

The ratio $R(s) = \sigma(e^+e^- \to hadrons)/\sigma(e^+e^- \to \mu^+\mu^-)$, characterizing the total cross section of $e^+e^-$ annihilation into hadrons, provides the simplest example of the analytic continuation of the effective QCD coupling constant $\alpha_s$ into the timelike region. The standard procedure (see, e.g., [27] and references cited therein) is to calculate the Adler function $D(Q^2)$ by taking the derivative $D(Q^2) = Q^2 d\Pi/dQ^2$ of the vacuum polarization $\Pi(Q^2)$ related to $R(s)$ by

$$R(s) = \frac{1}{2\pi i} \left( \Pi(-s + i\epsilon) - \Pi(-s - i\epsilon) \right) . \qquad (2)$$

In perturbative QCD, $D(Q^2)$ is given by the $\alpha_s(Q^2)$-expansion:

$$D^{\text{QCD}}(Q^2) = \sum_q e_q^2 \left\{ 1 + \frac{\alpha_s(Q^2)}{\pi} + d_2 \left( \frac{\alpha_s(Q^2)}{\pi} \right)^2 + d_3 \left( \frac{\alpha_s(Q^2)}{\pi} \right)^3 + \ldots \right\} . \qquad (3)$$

In the $\overline{\text{MS}}$ scheme, the coefficients $d_i$ are known up to $i = 3$ [27,28]. Using Eq. (2) and the definition of $D$, one can relate $R^{\text{QCD}}(s)$, the perturbative QCD version of $R(s)$, directly to $D^{\text{QCD}}(Q^2)$

$$R^{\mathrm{QCD}}(s) = \frac{1}{2\pi i} \int_{-s-i\epsilon}^{-s+i\epsilon} D^{\mathrm{QCD}}(\sigma) \frac{d\sigma}{\sigma} \ . \qquad (4)$$

The integration contour in Eq.(4) goes below the real axis from $-s - i\epsilon$ to some point $Q^2$ in the deep spacelike region and then above the real axis to $-s + i\epsilon$, i.e. in the region where the function $D(s)$ is analytic.

In a shorthand notation, $D \to R \equiv \Phi[D]$. The actual calculation is very simple if one represents $\alpha_s(Q^2)$ through an expansion in $1/\ln(Q^2/\Lambda^2)$, i.e., via the $\Lambda$-parameterization. The latter results from the QCD Gell-Mann-Low equation

$$L \equiv \ln(Q^2/\Lambda^2) = \frac{4\pi}{b_0 \alpha_s} + \frac{b_1}{b_0^2} \ln\left(\frac{\alpha_s}{4\pi}\right) + \Delta + \frac{b_2 b_0 - b_1^2}{b_0^3} \frac{\alpha_s}{4\pi} + O(\alpha_s^2) \ , \qquad (5)$$

where $b_k$ are $\beta$-function coefficients:

$b_0 = 11 - 2N_f/3$ [1], $b_1 = 102 - 38N_f/3$ [29], $b_2^{MS} = 2857/2 - 5033N_f/18 + 325N_f^2/54$ [30].

Inverting (5) by iterations and reexpanding the result in $1/L$ we get the $\Lambda$-parameterization for the running coupling constant

$$\alpha_s(Q^2) = \frac{4\pi}{b_0 L} \left\{ 1 - \frac{L_1}{L} + \frac{1}{L^2} \left[ L_1^2 - \frac{b_1}{b_0^2} L_1 + \frac{b_2 b_0 - b_1^2}{b_0^4} \right] + O(1/L^3) \right\} \ , \qquad (6)$$

where $L_1 = (b_1/b_0^2) \ln(b_0 L) - \Delta$ [31,32]. To fix the functional dependence of $\alpha_s(Q^2)$ on $Q^2$, one should specify the integration constant $\Delta$. The standard (or "popular") choice is

$$\Delta^{pop} = \frac{b_1}{b_0^2} \ln b_0 \qquad (7)$$

which gives the shortest expression $(b_1/b_0^2) \ln(L)$ for $L_1$. A clear disadvantage of this choice is that it guarantees a rather large $1/L^2$ correction to $\alpha_s$, which results in a large difference between $\Lambda^{LO}$ and $\Lambda^{NLO}$. As argued in Ref. [13], a more appropriate (optimal) choice is

$$\Delta^{opt} = \frac{b_1}{b_0^2} \ln b_0 \bar{L} \ , \qquad (8)$$

where $\bar{L}$ is the average value of the logarithm $L$ within the region under study, e.g., $\bar{L} = 4$ corresponding to $\alpha_s/\pi \sim 0.1$. For this choice, the ratio $L_1/L$ is smaller than 7% and Eq.(6) has 1% accuracy in the whole region $L > 3$, with the total correction to the simplest formula (1) being less than 10%.

The $\Lambda$-parameters corresponding to different $\Delta$'s are related by

$$\Lambda_2 = \Lambda_1 e^{(\Delta_1 - \Delta_2)/2} \ . \qquad (9)$$

In particular,

$$\Lambda^{opt} = \Lambda^{pop}/\bar{L}^{b_1/2b_0^2} \ . \qquad (10)$$

Taking $\bar{L} = 4$ we get $\Lambda^{opt}|_{\bar{L}=4} \approx \Lambda^{pop}/1.73$. In connection with the discussion above, we want to stress here that preparing to analytically continue an approximate expression it makes sense to take care of the convergence quality of the original expansion in the spacelike region. If there are corrections which are under our full control and we can make them small, then we should use this opportunity and make them small.

Now one can substitute $\alpha_s(Q^2)$ in Eq.(3) by its $\Lambda$-parameterization to get an $1/L$ expansion for the Adler function $D(Q^2)$. For each term of this expansion, the integral (4) can be calculated explicitly (see also [33])

$$1 \to 1 \ , \qquad (11)$$

$$\frac{1}{L_\sigma} \to \frac{1}{\pi}(\pi/2 - \arctan(L_s/\pi))\bigg|_{s>\Lambda^2} = \frac{1}{\pi}\arctan(\pi/L_s) = \frac{1}{L_s}\left\{1 - \frac{1}{3}\frac{\pi^2}{L_s^2} + \ldots\right\} \ , \qquad (12)$$

$$\frac{\ln(L_\sigma/L_0)}{L_\sigma^2} \to \frac{\ln(\sqrt{L_s^2 + \pi^2}/L_0) - (L_s/\pi)(\pi/2 - \arctan(L_s/\pi)) + 1}{L_s^2 + \pi^2}\bigg|_{s>\Lambda^2}$$

$$= \frac{\ln(\sqrt{L_s^2 + \pi^2}/L_0) - (L_s/\pi)\arctan(\pi/L_s) + 1}{L_s^2 + \pi^2} = \frac{L_s/L_0}{L_s^2}\left\{1 - \frac{\pi^2}{L_s^2} + \ldots\right\} + \frac{5}{6}\frac{\pi^2}{L_s^4} + \ldots \ , \qquad (13)$$

$$\frac{1}{L_\sigma^2} \to \frac{1}{L_s^2 + \pi^2} = \frac{1}{L_s^2}\left\{1 - \frac{\pi^2}{L_s^2} + \dots\right\} , \tag{14}$$

$$\frac{1}{L_\sigma^n} \to (-1)^n \frac{1}{(n-1)!}\left(\frac{d}{dL_s}\right)^{n-2} \frac{1}{L_s^2+\pi^2} = \frac{1}{L_s^n}\left\{1 - \frac{\pi^2}{L_s^2}\frac{n(n+1)}{6} + \dots\right\} , \tag{15}$$

where $L_s = \ln(s/\Lambda^2)$, $L_\sigma = \ln(\sigma/\Lambda^2)$, and we assume that $s > 0$. Furthermore, $L_0 = e^{\Delta b_0^2/b_1}/b_0$ is a constant depending on the $\Delta$-choice in the $\Lambda$-parameterization.

Using (6) and incorporating Eqs.(11)-(15) (as well as their generalizations for $\ln^2 L/L^3, \ln L/L^3$, etc.) one obtains the expansion for $R^{\rm QCD}(s)$

$$R^{\rm QCD}(s) = \sum_q e_q^2 \left\{1 + \sum_{k=1} d_k \Phi[(\alpha_s/\pi)^k]\right\} \tag{16}$$

in which all the $(\pi^2/L^2)^N$-terms are resummed.

As noted in Ref. [13], the application of the $\Phi$-operation normally violates nonlinear relations: $\Phi[1/L^2] \neq (\Phi[1/L])^2$, etc. However, it respects linear relations $\Phi[A+B] = \Phi[A] + \Phi[B]$, $\Phi[\lambda A] = \lambda \Phi[A]$ and

$$\Phi\left[\frac{dD}{dL_\sigma}\right] = \frac{d}{dL_s}\Phi[D] . \tag{17}$$

In particular, this relation was used to explicitly obtain $\Phi[1/L^n]$ in Eq.(15). As a result, expansion (16) is not an expansion in powers of some particular parameter. A priori, there is no reason to believe that a power series expansion is better than any other. In fact, expansion (16) converges better than the generating expansion (4) for $D(\sigma)$ because, as it follows from Eqs. (12)-(15), $\Phi[\alpha_s^N]$ is always smaller than $\alpha_s^N$. Moreover, $(\Phi[\alpha_s^{N+1}])^{1/(N+1)} < (\Phi[\alpha_s^N])^{1/N}$, i.e., the effective expansion parameter decreases in higher orders. Thus, if one succeeded in obtaining a good $\alpha_s^N$ expansion for $D(\sigma)$ (with all $d_N$ being small numbers), then the resulting $\Phi[\alpha_s^N]$ expansion for $R^{\rm QCD}(s)$ is even better, and the best thing to do is to leave it as it is.

The timelike analogue of the simplest $\Lambda$-parameterization for $\alpha_s(Q^2)$ (Eq.(1)) is then

$$\tilde{\alpha}_s(q^2) = \frac{4}{b_0}\left[\frac{\pi}{2} - \arctan\left(\frac{\ln(q^2/\Lambda^2)}{\pi}\right)\right]\bigg|_{s>\Lambda^2} = \frac{4}{b_0}\arctan\left(\frac{\pi}{\ln(q^2/\Lambda^2)}\right) . \tag{18}$$

This function has a finite value both at $q^2 = \Lambda^2$ and $q^2 = 0$.

The well-known deficiency of the perturbative expansion for $D^{\rm QCD}(Q^2)$ in powers of $\alpha_s(Q^2)$ is the presence of the unphysical singularity at $Q^2 = \Lambda^2$ induced by the Landau pole of $1/\ln(Q^2/\Lambda^2)$. As a consequence, $R^{\rm QCD}(s)$ as calculated from Eq. (4), also has unphysical features: namely, it does not vanish on the negative real axis. In particular, substituting $1/L_\sigma$ into the integral (4) and taking negative $s$ we get

$$\frac{1}{L_\sigma}\bigg|_{s<0} \to \theta(-\Lambda^2 \le s \le 0) , \tag{19}$$

which results in an unphysical cut of $\Pi^{\rm QCD}(s)$ in the region $-\Lambda^2 \le s \le 0$. Furthermore, applying Eq. (4) to the pole term $D^{\rm pole}(Q^2) = \Lambda^2/(Q^2 - \Lambda^2)$ one obtains the result coinciding with the rhs of Eq.(19). Hence, if one now postulates that $D^{\rm QCD}(Q^2)$ is given by integrating $R^{\rm QCD}(s)$ over the physical region $s > 0$ only, i.e., if one takes

$$\tilde{D}^{\rm QCD}(Q^2) = Q^2 \int_0^\infty \frac{R^{\rm QCD}(s)}{(s+Q^2)^2} ds \tag{20}$$

(this transformation will be denoted as $R \to \tilde{D}$), then $\tilde{D}^{\rm QCD}(Q^2)$ is free from the unphysical singularities at $Q^2 = \Lambda^2$. For instance, combining the two transformations $(D \to R \to \tilde{D}) \equiv (D \Rightarrow \tilde{D})$ one would get

$$\frac{4\pi}{b_0 \ln(Q^2/\Lambda^2)} \Rightarrow \frac{4\pi}{b_0}\left(\frac{1}{\ln(Q^2/\Lambda^2)} - \frac{\Lambda^2}{Q^2 - \Lambda^2}\right) \equiv \bar{\alpha}_s(Q^2), \tag{21}$$

which coincides with the pole-free expression for the running coupling constant proposed by Shirkov and Solovtsov [34]. However, since the $D \to R$ operation does not respect nonlinear relations, the $D \Rightarrow \tilde{D}$ transformation acting on $1/L_\sigma^n$ would not produce the $n$th power of the rhs of Eq.(21). Hence, $\bar{\alpha}_s$ cannot serve as an expansion parameter of a power series. Noting that both $D \to R$ and $R \to \tilde{D}$ convert derivatives with respect to the logarithm of the initial variable into derivatives with respect to the logarithm of the resulting variable we obtain

$$\frac{1}{L_{Q^2}^n} = (-1)^n \frac{1}{(n-1)!} \frac{d^{n-1}}{L_{Q^2}^{n-1}} \frac{1}{L_{Q^2}} \Rightarrow (-1)^n \frac{1}{(n-1)!} \frac{d^{n-1}}{dL_{Q^2}^{n-1}} \left( \frac{1}{L_{Q^2}} - \frac{\Lambda^2}{Q^2 - \Lambda^2} \right). \qquad (22)$$

This relation was given in a recent paper by Shirkov [35], see also Ref. [36] for a related discussion of perturbation theory expansions in the timelike and spacelike regions.

For moderate values of $Q^2$, the modification due to the continuation into the timelike region is numerically rather significant: for $\alpha_s \gtrsim 0.3$ the $\pi^2/L^2$-terms change $\alpha_s$ by more than 20%, i.e., they are more important (for an optimal choice of the $\Delta$-parameter) than the 2-loop corrections in the $\Lambda$-parameterization (6). On the other hand, the difference between $\tilde{\alpha}_s(Q^2)$ and the modified spacelike coupling $\bar{\alpha}_s(Q^2)$ (taken at mirror momenta) is rather small (less than 10%) for all values of $Q^2$.

Thus, using the $\Lambda$-parameterization for the effective QCD coupling constant in the spacelike region, we obtained an explicit expansion for the timelike quantity $R^{\mathrm{QCD}}(s)$. One may question, though, the reliability of the above formulas in the region of small momenta $|q| \sim \Lambda$. In particular, a rapid change of $\tilde{\alpha}_s$ in the small-$q^2$ region (compare $\tilde{\alpha}_s(\Lambda^2) = 2\pi/b_0$ and $\tilde{\alpha}_s(0) = 4\pi/b_0$) is as suspicious as the Landau pole of $\alpha_s(Q^2)$. Evidently, they both are artifacts of the analytic continuation procedure applied outside the applicability region. It is well known that the physical $R(s)$ vanishes below the two-pion threshold and approaches the perturbative value only for values of $s$ marginally larger than $\Lambda^2$. So, one may argue that a more realistic procedure is to integrate $R^{QCD}(s)$ in the dispersion relation (20) from some effective threshold $s_0$ rather than from zero. Taking, e.g., $s_0 = \Lambda^2$, one would get another effective spacelike coupling, call it $\hat{\alpha}_s(Q^2)$. It vanishes at $Q^2 = 0$, but is essentially constant $\hat{\alpha}_s(Q^2)/\pi \approx 0.1$ in a wide range $\Lambda^2 \lesssim Q^2 \lesssim 30\Lambda^2$ of momenta. Hence, $\hat{\alpha}_s(Q^2)$ effectively "freezes" at small momenta (see also [37–39,33]).

## III. COLLINEAR LOGARITHMS AND DISTRIBUTION AMPLITUDES IN THE TIMELIKE REGION

The logarithmic dependence on the large momentum scale $Q^2$ may also appear through mass logarithms $\ln(Q^2/m^2)$, where $m$ is some mass or an infrared regularization parameter. Note that the standard pQCD factorization

$$T(Q^2/m^2) = t(Q^2/\mu^2) \otimes \varphi(\mu^2) \qquad (23)$$

works only in a single-logarithm situation, when there may appear just one $\ln(Q^2/m^2)$ factor per each loop. These collinear logarithms can be absorbed into the renormalization of the long-distance function (distribution amplitude) $\varphi(\mu^2)$. In particular, taking $\mu^2 = Q^2$, one arrives at the description in terms of $Q^2$-dependent functions $\varphi(Q^2)$. Again, if the large momentum is timelike, the collinear logarithms $\ln(Q^2/m^2)$ acquire the imaginary part $\pm i\pi$, and we may ask how one should define the $Q^2$-dependent distribution amplitudes $\varphi(Q^2)$ in the timelike region.

To approach this problem, let us consider the simplest example of a hard exclusive process: $\pi^0$ production in $\gamma^*\gamma$ collisions. Its pQCD expansion starts at zero order in $\alpha_s$

$$t_0(x, Q^2) = \frac{1}{xQ^2}, \qquad (24)$$

and the leading pQCD result [40] for the large-$Q^2$ behavior of the form factor is

$$F_{\gamma^*\gamma\pi}(Q^2) = \frac{4\pi}{3} \int_0^1 \frac{\varphi_\pi(x)}{xQ^2} dx \equiv \frac{4\pi f_\pi}{3Q^2} I_0. \qquad (25)$$

The nonperturbative information here is accumulated in the same integral

$$I_0 = \frac{1}{f_\pi} \int_0^1 \frac{\varphi_\pi(x)}{x} dx \qquad (26)$$

that appears in the one-gluon-exchange diagram for the pion electromagnetic form factor [41–43]. The value of $I$ depends on the shape of the pion distribution amplitude $\varphi_\pi(x)$. In particular, using the asymptotic form [41,42]

$$\varphi_\pi^{\mathrm{as}}(x) = 6 f_\pi x(1-x) \qquad (27)$$

gives $I_0^{\mathrm{as}} = 3$. If one takes instead the Chernyak–Zhitnitsky model [44]

$$\varphi_\pi^{CZ}(x) = 30 f_\pi x(1-x)(1-2x)^2, \tag{28}$$

the integral $I_0$ increases by a sizable factor of 5/3: $I_0^{CZ} = 5$. This difference can be used for an experimental discrimination between the two competing models for the pion distribution amplitude.

At one loop, the $\overline{\text{MS}}$ coefficient function for the $\gamma^*\gamma \to \pi^0$ form factor was calculated in refs. [45–47] and was found to be

$$t(x, Q^2; \mu^2) = \frac{1}{xQ^2}\left\{1 + C_F \frac{\alpha_s}{2\pi}\left[\left(\frac{3}{2} + \ln x\right)\ln(Q^2/\mu^2) + \frac{1}{2}\ln^2 x - \frac{x \ln x}{2(1-x)} - \frac{9}{2}\right]\right\}. \tag{29}$$

In full compliance with the factorization theorems [48,40] (see also [49–51]), the one-loop contribution contains no Sudakov double logarithms $\ln^2 Q^2$ of the large momentum transfer $Q$. Physically, this result is due to the color neutrality of the pion. In the axial gauge, the Sudakov double logarithms appear in the box diagram but they are canceled by similar terms from the quark self-energy corrections. In Feynman gauge, the double logarithms $\ln^2 Q^2$ do not appear in any one-loop diagram. It is easy to check that the term containing the logarithm $\ln(Q^2/\mu^2)$ has the form of a convolution

$$\frac{1}{xQ^2} C_F \frac{\alpha_s}{2\pi}\left(\frac{3}{2} + \ln x\right) = \int_0^1 \frac{1}{\xi Q^2} V(\xi, x)\, d\xi \tag{30}$$

of the lowest-order ("Born") term $t_0(\xi, Q^2) = 1/\xi Q^2$ and the kernel

$$V(\xi, x) = \frac{\alpha_s}{2\pi} C_F \left[\frac{\xi}{x}\theta(\xi < x)\left(1 + \frac{1}{x-\xi}\right) + \frac{\bar{\xi}}{\bar{x}}\theta(\xi > x)\left(1 + \frac{1}{\xi-x}\right)\right]_+ \tag{31}$$

governing the evolution of the pion distribution amplitude. The "+"-operation is defined here, as usual [52], by

$$[F(\xi, x)]_+ = F(\xi, x) - \delta(\xi - x)\int_0^1 F(\zeta, x)\, d\zeta. \tag{32}$$

Since the asymptotic distribution amplitude is the eigenfunction of the evolution kernel $V(\xi, x)$ corresponding to zero eigenvalue

$$\int_0^1 V(\xi, x)\, \varphi^{\text{as}}(x)\, dx = 0, \tag{33}$$

the coefficient $(\frac{3}{2} + \ln x)$ of the $\ln(Q^2/\mu^2)$ term vanishes after the $x$-integration with $\varphi^{\text{as}}(x)$. Hence, the size of the one-loop correction for the asymptotic distribution amplitude is $\mu$-independent and is determined only by the remaining terms (for a detailed discussion of their structure, see Ref. [53]).

In this section, we want to concentrate on the $Q^2$-dependence induced by collinear logarithms, which in this process start to appear at the one-loop level. The UV logarithms shifting the argument of $\alpha_s$ appear only at two-loop order. Hence, analyzing the leading collinear logarithms $(\alpha_s \ln(Q^2/\mu^2))^N$ we will treat $\alpha_s$ as a constant. The factorization theorem means essentially that the leading logarithms $(\alpha_s \ln(Q^2/\mu^2))^N$ exponentiate in higher orders producing a factor which can be absorbed into the renormalization of the pion distribution amplitude

$$\varphi(\mu^2) \to \exp[-\ln(Q^2/\mu^2)\, V] \otimes \varphi(\mu^2). \tag{34}$$

Now, taking a timelike momentum $Q^2 = -q^2$, we would get an extra $\pm i\pi$ term: $\ln(Q^2/\mu^2) \to \ln(q^2/\mu^2) \pm i\pi$ and

$$\exp[-\ln(Q^2/\mu^2)\, V] \to \exp[-\ln(q^2/\mu^2)\, V]\exp[\pm i\pi V]. \tag{35}$$

The first exponential corresponds to the standard evolution of the pion distribution amplitude from the scale $\mu^2$ to the scale $q^2$. The second exponential is specific for the timelike kinematics. In our approximation, it is $q^2$-independent and can be treated as a conversion factor for the transition from a "spacelike" distribution amplitude $\varphi$ to its timelike counterparts $\tilde{\varphi}_\pm$

$$\tilde{\varphi}_\pm = \exp[\pm i\pi V] \otimes \varphi. \tag{36}$$

In general, "timelike" distribution amplitudes have both real and imaginary parts. However, since $V \otimes \varphi^{\text{as}} = 0$, the spacelike asymptotic distribution amplitude does not differ from its timelike counterpart.

To estimate the effect of phases, let us consider the case when the spacelike distribution amplitude is given by the Chernyak-Zhitnitsky (CZ) model [44], which can be represented as

$$\varphi^{CZ} = \varphi^{as} + \varphi_2 , \qquad (37)$$

where $\varphi^{as} = 6f_\pi x(1-x)$ and $\varphi_2 = 24 f_\pi x(1-x)(1-5x(1-x))$ is the next eigenfunction of the $V$ kernel corresponding to the eigenvalue $\gamma_2 = \frac{25}{18}\alpha_s/\pi$. The timelike distribution amplitude is then

$$\tilde{\varphi}^{CZ}_\pm = \varphi^{as} + e^{\pm i(25/18)\alpha_s}\varphi_2 \qquad (38)$$

and the $I$ integral for this function is

$$\tilde{I}^{CZ}_\pm = 3 + 2e^{\pm i(25/18)\alpha_s} . \qquad (39)$$

Its absolute magnitude

$$\left|\tilde{I}^{CZ}\right| = 5\sqrt{1 - \frac{24}{25}\sin^2\left(\frac{25}{36}\alpha_s\right)} \qquad (40)$$

is slightly smaller (by 2% if $\alpha_s = 0.3$) than the spacelike value $I^{CZ} = 5$.

## IV. SUDAKOV LOGARITHMS AND $K$-FACTOR

Small radiative corrections in the timelike version of the $\gamma^*\gamma \to \pi^0$ process are in strong contrast with the large $K$-factor value found for the Drell–Yan process $AB \to \gamma^* X$. These corrections originate from the Sudakov double logarithms $(\alpha_s \ln^2(Q^2/m^2))^N$. In the spacelike region, the double logarithms due to the virtual gluon exchanges exponentiate into the Sudakov form factor

$$S(Q^2/m^2) = e^{-\alpha_s \ln^2(Q^2/m^2)/3\pi} \qquad (41)$$

(again, we treat $\alpha_s$ as a constant). In the DY process, the photon momentum is timelike, and the logarithm $\ln(Q^2/m^2)$ acquires the $\pm i\pi$ additional term, so that one has

$$-L^2 \to -L^2 \pm 2i\pi L + \pi^2 . \qquad (42)$$

The imaginary parts of the two conjugate diagrams shown in Fig.1a,b cancel, the double log $L^2$ from Fig.1a (b) is also canceled by the real gluon emission diagram Fig.1c (d), while the $\pi^2$-term survives and leads, after exponentiation, to a large $K$ factor $\exp[2\pi\alpha_s/3] \sim 2$. The crucial technical observation here is that the real emission diagrams give $L^2$ without $\pi^2$-terms. This can be easily understood looking at the reduced diagrams for the virtual vertex correction and real gluon emission. Take for definiteness, the Feynman gauge. Then the virtual vertex correction diagram Fig.1a contains the $-\ln^2(-s/m^2)$ term, where $s = (xp_A + yp_B)^2 = xyS$ is timelike, and the resulting contribution contains a $\pi^2$ term. The real emission diagram Fig.1c, in turn, contains the $\ln^2(-u/m^2)$ term, where $u = (xp_A - yp_B)^2 = -xyS$ is now spacelike, and there is no $\pi^2$ term in this contribution.

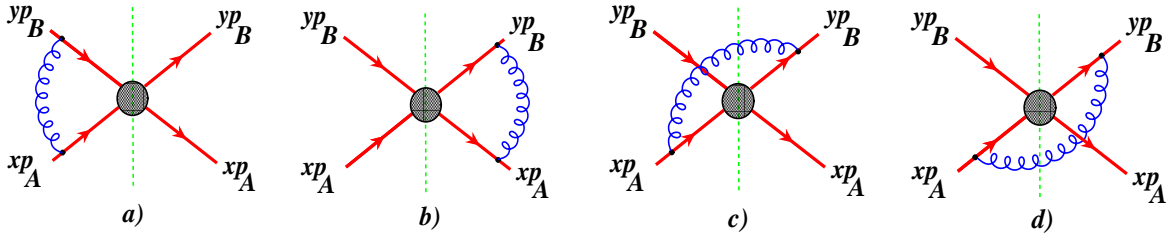

FIG. 1. One-loop reduced diagrams for the DY process cross section calculated as imaginary part of the forward scattering amplitude. $a, b$) Virtual vertex corrections. $c, d$) Real gluon emission corrections.

For the hard pQCD contribution to the pion electromagnetic form factor (which is considered in more detail in the next section), the situation is completely different. In this case, the initial state of the hard subprocess is represented by a $q\bar{q}$ pair with momenta $xp$ and $(1-x)p$. After the hard scattering subprocess, one deals with a $q\bar{q}$ pair with the final momentum $p'$ shared in fractions $yp'$ and $(1-y)p'$. In Feynman gauge, the double logarithms $\ln^2(Q^2/\mu^2)$, where $Q^2 = -(p-p')^2$, appear when the reduced diagrams have the structure of those shown in Fig.2. One can easily check that the relevant momentum transfers in all four cases have the structure $t_{ij} = (x_i p - y_j p')^2 = -x_i y_j Q^2$, resulting in the double logs $\ln^2(-t_{ij}/\mu^2)$. When the momentum transfer $q = p' - p$ is spacelike, all $t_{ij}$'s are spacelike, whereas for a timelike $q$, all $t_{ij}$'s are timelike as well. In the latter case, one has $\pi^2$ terms for each particular diagram. The double logarithms in diagrams 2a and 2b (2c and 2d) differ in sign because the soft gluon interacts in the final state with quarks of opposite color charge. Hence, due to the color neutrality of the pion, the double logs $\ln^2(Q^2/\mu^2)$ cancel for the sum of the diagrams of a given order. For timelike $q$, they cancel together with the accompanying $\pi^2$ terms.

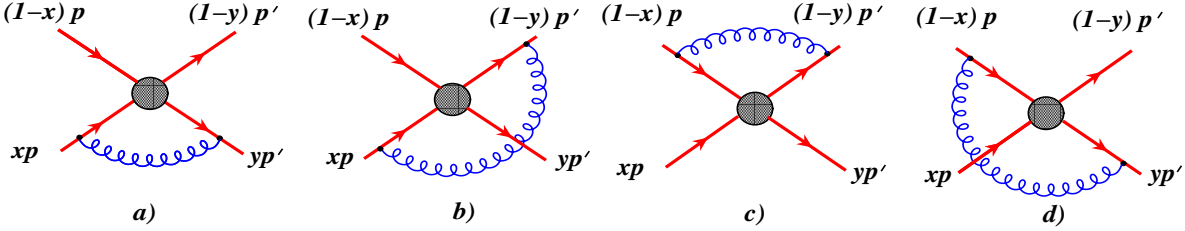

FIG. 2. One-loop reduced diagrams for the hard pQCD contribution to the pion electromagnetic form factor, which contain the Sudakov double logarithms in Feynman gauge.

Thus, even for a timelike momentum transfer $q$, there is no $K$-factor for the pQCD hard contribution to the pion electromagnetic form factor. After cancellation of the Sudakov double logs, only the evolution-related collinear logarithms remain, and the situation is rather similar to the simplest case of the $\gamma^*\gamma \to \pi^0$ form factor.

## V. PION FORM FACTOR IN THE PERTURBATIVE QCD APPROACH

The general pQCD factorization formula for the pion electromagnetic form factor at large momentum transfer reads

$$F_\pi^{\text{HARD}}(Q^2) = \int_0^1 dx \int_0^1 \varphi(x,\mu_F^2,\mu_R^2)\, T(x,y,Q^2,\mu_F^2,\mu_R^2)\, \varphi(y,\mu_F^2,\mu_R^2)\, dy\;, \tag{43}$$

where $\mu_F$ is the factorization scale for the collinear logarithms and $\mu_R$ is the renormalization scale for the UV logarithms. The hard scattering amplitude is given by an expansion in $\alpha_s$

$$T(x,y,Q^2) = \frac{2\pi C_F \alpha_s(\mu_R^2)}{xyQ^2 N_c}\left[1 + \frac{\alpha_s(\mu_R^2)}{2\pi} T^{(1)}(x,y,Q^2,\mu_F^2,\mu_R^2) + O(\alpha_s^2)\right]\;. \tag{44}$$

The one-loop correction $T^{(1)}(x,y,Q^2,\mu_F^2,\mu_R^2)$ was calculated using the dimensional regularization in several papers [54–59] which differ from each other by a particular choice of renormalization and factorization prescriptions. These differences (and also typos and mistakes) were discussed in Refs. [58–60,39]. In the $\overline{MS}$ subtraction scheme, supplemented by the requirement that both the $\alpha_s$ and $\varphi_\pi(x)$ are process-independent functions, the one-loop correction has the form

$$\begin{aligned}
T^{(1)} &= C_F T^F(x,y,Q^2,\mu_F^2) + \frac{b_0}{2} T^\beta(x,y,Q^2,\mu_R^2) + (C_F - N_c/2)\, T^A(x,y)\;, \\
T^F &= \left[3 + \ln(xy)\right] \ln\left(\frac{Q^2}{\mu_F^2}\right) + \frac{1}{2}\ln^2(xy) + \frac{5}{2}\ln(xy) - \frac{x\ln x}{2(1-x)} - \frac{y\ln y}{2(1-y)} - \frac{14}{3}\;, \\
T^\beta &= -\ln\left(\frac{Q^2}{\mu_R^2}\right) - \ln(xy) + \frac{5}{3}\;, \\
T^A &= \mathbf{Li}_2(1-x) - \mathbf{Li}_2(x) + \ln(1-x)\ln\left(\frac{y}{1-y}\right) - \frac{5}{3} \\
&\quad + \frac{1}{(x-y)^2}\left((x+y-2xy)\ln(1-x) + 2xy\ln(x) + \frac{(1-x)x^2 + (1-y)y^2}{x-y}\right.
\end{aligned} \tag{45}$$

$$\times \left[ \ln(1-x)\ln(y) - \mathbf{Li}_2(1-x) + \mathbf{Li}_2(x) \right] \right) + \{x \leftrightarrow y\} \tag{46}$$

(we use here notations similar to those of Ref. [61]). As usual, $\mathbf{Li}_2$ is the dilogarithm (Spence) function.

We already discussed in the previous section that all the Sudakov double logarithms $\ln^2(Q^2/\mu^2)$ cancel and that only the collinear single-logarithms $\ln(Q^2/\mu_F^2)$ remain. Comparing Eq. (46) with the one loop correction to the $\gamma^*\gamma \to \pi^0$ hard scattering amplitude, Eq. (29), one can easily notice many similarities in the structure of the one-loop corrections in these two cases. In particular, the coefficient $[3 + \ln(xy)]$ in front of the evolution logarithm $\ln(Q^2/\mu_F^2)$ is the sum $[3/2 + \ln x] + [3/2 + \ln y]$ of terms corresponding to the convolution of the tree level term $1/\xi\eta$ with the kernels $V(x,\xi)\delta(\eta - y)$ and $\delta(x - \xi)V(\eta, y)$ (see Eq. (30) ) inducing the evolution of the pion distribution amplitudes $\varphi(x,\mu_F^2)$ and $\varphi(y,\mu_F^2)$. In a sense, the collinear logarithms indicate that the pion structure is probed at a scale proportional to $Q$. However, one should remember that since the asymptotic wave function does not evolve, the coefficient accompanying the evolution logarithm $\ln(Q^2/\mu_F^2)$ vanishes if the pion wave function has the asymptotic shape. As a result, the choice of $\mu_F$ in that case does not affect the size of the one-loop correction. The latter comes from several sources which can be identified in a way similar to the detailed analysis of the one loop correction for the $\gamma^*\gamma \to \pi^0$ form factor given in Ref. [53].

In addition to the evolution term proportional to $\ln(Q^2/\mu_F^2)$, there is a rather large positive correction due to the $\frac{1}{2}\ln^2(xy)$ term and even larger negative contributions corresponding to the constant term $-14/3$ and the logarithmic term $\frac{5}{2}\ln(xy)$. As explained in Ref. [53], in the $\gamma^*\gamma \to \pi^0$ case, the $\frac{1}{2}\ln^2 x$ term is a result of a positive $\ln^2 x$ evolution-related contribution and a negative $-\frac{1}{2}\ln^2 x$ Sudakov-related term. As we emphasized earlier, the Sudakov $\ln^2 Q^2$ double logs should cancel, otherwise there is no pQCD factorization. However, when several scales are involved, like $Q^2$ and $xyQ^2$ in our case, there may be a remnant like $\ln^2(xy)$. In the pion form factor case, there is another scale $xQ^2$, the quark virtuality, whence the single logarithms $\ln x + \ln y$. The latter give a rather large negative contribution. There are also large negative constants ($-9/2$ in the $\gamma^*\gamma \to \pi^0$ case and $-14/3$ in the pion form factor case), which are another (and numerically very important) manifestation of the Sudakov effects in the impact parameter space. In full analogy with the results of Ref. [53], these [and the $\ln^2(xy), \ln(xy)$ terms] result from convoluting the $b$-space version $K_0(\sqrt{xyQ^2b^2})$ of the one-gluon exchange propagator and the $b$-space Sudakov form factors $S(x, bQ)$, $S(y, bQ)$ (exact one-loop expressions are given in [53]). In the practically important case of the asymptotic wave function, the total correction due to the $T^F$ term is negative and equal to $-(71/18)\alpha_s/\pi$; as one could expect, it is approximately twice larger than that in the $\gamma^*\gamma \to \pi^0$ case.

The situation is reversed in the case of the UV related $T^\beta$ term: it is dominated by large positive contributions. In full accordance with the renormalization group, the UV logarithm $\ln(Q^2/\mu_R^2)$ is accompanied by the $\beta$-function coefficient $b_0$. It generates the running of the effective QCD coupling $\alpha_s$, "suggesting" that we should use some scale proportional to $Q^2$ as its argument. According to Brodsky, Lepage and Mackenzie [62], one should choose the argument of the effective coupling constant in such a way as to absorb all the terms proportional to $b_0$ from the next loop correction. Taken literally, the BLM prescription in our case corresponds to using the (rescaled) gluon virtuality

$$\mu_R^2 = xyQ^2 e^{-5/3} \approx xy(Q/2.3)^2 \tag{47}$$

as the argument of $\alpha_s$. The rescaling factor $e^{5/6} \approx 2.3$ reflects the fact that the $\overline{MS}$ scheme measures the momenta in "wrong" units. To cure this effect, one may introduce a version of the minimal subtraction scheme which measures momenta in more "physical" units $\Lambda_{\text{PHYS}} = e^{5/6}\Lambda_{\overline{MS}}$. This choice is similar to using the $\alpha_V$ coupling of Brodsky et al. [39] (note that their relation $\alpha_V(Q) = \alpha_s^{\overline{MS}}(e^{-5/6}Q)[1 - \frac{2}{3}N_c\alpha_s^{\overline{MS}}/\pi]$ includes a NLO correction). One should remember, however, that the actual expansion parameter for switching from the leading to the next-to-leading level is $1/\ln(Q^2/\Lambda^2)$ rather than $\alpha_s$ as a whole. As a result, the "non-physical" nature of the $\overline{MS}$ scheme is almost totally compensated by the non-optimal "popular" choice for the analytic form of $\alpha_s(Q^2)$. As discussed in Section II, $\Lambda^{opt} \approx \Lambda^{pop}/1.74$. As a result, $\Lambda_{\text{PHYS}}^{opt} \approx 1.3\Lambda_{\overline{MS}}^{pop}$. Due to the compensation of two opposite corrections, the standard $\Lambda_{\overline{MS}}^{pop}$ parameter is rather close to the genuine "$\Lambda_{\text{QCD}}$" parameter of the $\text{PHYS}^{opt}$ scheme in which the coupling $\alpha_s(k^2)$ corresponding to the vertex with the gluon momentum $k$ is given by $4\pi/b_0 \ln(k^2/\Lambda^2)$ without sizable next-to-leading order corrections. In other words, using the NLO expression for $\alpha_s$ in the popular form is equivalent to adding a negative term $-(b_1/b_0^2)\ln L$ to $T^\beta$, partially compensating the "5/3" constant. For $L \sim 4$, this reduces 5/3 by a factor of 3. Choosing "PHYS" vs. $\overline{MS}$ and "opt" vs. "pop" one reduces both types of corrections which iterate in higher orders. As stated earlier, if the size of some corrections is under our control, it is preferable to keep them small rather than rely on cancellation of large terms. The closeness of $\Lambda_{\text{PHYS}}^{opt}$ to $\Lambda_{\overline{MS}}^{pop}$ means that discussing the pQCD applicability region one should compare the $\Lambda^2$ parameter of the $\overline{MS}^{pop}$ scheme with the actual (unrescaled) gluon virtuality $xyQ^2$.

However, taking the argument of the effective coupling constant proportional to $xyQ^2$ one faces the following problem: since the integration is over all the momentum fractions in the range $0 \leq x, y \leq 1$, the "short-distance" amplitude in this case always gets contributions from the infrared region of arbitrarily small virtualities. In this sense, such an "inside the integral" BLM prescription contradicts the spirit of the pQCD factorization ideology which aims at a perfect separation of the short-distance and long-distance effects (at least in perturbation theory). The consistent pQCD approach is to apply the BLM prescription to the form factor as a whole, i.e., "outside the integral". In this case, one should choose $\mu_R$ from the requirement that one should get zero for

$$\left\langle -\ln(Q^2/\mu_R^2) - \ln(xy) + \frac{5}{3} \right\rangle ,$$

where the "averaging" procedure $\langle \ldots \rangle$ stands for integration with $\varphi_\pi(x)\varphi_\pi(y)/xy$. This gives a universal $x,y$-independent scale $\mu_R = a_R Q$, which depends now on the shape of the distribution amplitude. For the asymptotic wave function, as we have seen, the average value of $\ln x$ is $-3/2$, hence the "outside the integral" BLM scale is (see also [39])

$$\mu_R^2|_{\varphi=\varphi^{\mathrm{as}}} = Q^2 \, e^{-5/3-3} \approx (e^{-5/3} Q^2)/20 . \tag{48}$$

As argued above, in the "pop" treatment, the factor $e^{-5/3}$ is largely compensated by the NLO corrections to $\alpha_s(Q^2)$, and, hence the essential virtuality of the "hard gluon" exchanged between the quarks is "only" by a factor of 20 smaller than $Q^2$, the nominal momentum transfer to the pion. Nevertheless, despite this sizable rescaling factor, the pQCD factorization approach is fully consistent in the asymptotic sense: for a sufficiently large $Q^2$ one can calculate the short-distance amplitude perturbatively in terms of an arbitrarily small expansion parameter $\alpha_s(Q^2/20)$. For comparison, in the case of Chernyak-Zhitnitsky wave function

$$\langle \ln x \rangle |_{\varphi^{CZ}} = -\frac{7}{3}$$

and $a_R^{CZ} \approx a_R^{\mathrm{as}}/2.3$: the essential gluon virtualities are 100 times smaller than $Q^2$. In this case, one should not expect early applicability of pQCD.

We would like to emphasize that the reason for such a drastic shift of the BLM scale to very low $\mu_R$ values is the positive large value of the $T^\beta$ correction: for $\mu_R = Q$ and $\varphi_\pi = \varphi_\pi^{\mathrm{as}}$, the $T^\beta$ term contributes the NLO correction $10\alpha_s/\pi$. One may be tempted to combine the large positive $T^\beta$ term and a sizable negative $T^F$ term to end up with a smaller total correction $\sim 6\alpha_s/\pi$. Physically, though, these corrections have a completely different nature: as argued above, the $T^F$ term comes primarily from the Sudakov effects. Since the latter exponentiate, one deals here with the $e^{-k_F \alpha_s}$ type series in which the sign of the corrections alternates. On the other hand, the UV corrections form a geometric series summed into $1/(1 - k_\beta \alpha_s)$. Hence, there is no doubt that a partial cancellation of the $\alpha_s$ terms will be followed by an amplified total correction at the $\alpha_s^2$ level. Leaving the physically unrelated Sudakov and UV corrections separate and addressing the region of experimentally accessible values of $Q^2 \lesssim 10\,\mathrm{GeV}^2$, one should take $\alpha_s$ at an infrared scale $\sim \Lambda$ (where it freezes at a value close to 0.3) and supplement the result by the exponential $\sim e^{-4\alpha_s/\pi} \sim 0.7$ of the negative one-loop corrections induced by the $b$-space Sudakov effects. Turning to the timelike momenta, we cannot find any sources of enhancement: for the $|k| \sim \Lambda$ region we see no other choice rather than to take the frozen value $\alpha_s \sim 0.3$ both in the spacelike and timelike regions, while for large momenta, the continuation of $\alpha_s(a_F^2 Q^2)$ converts $1/L$ into $1/(L + i\pi)$ and the ratio $|F_{\mathrm{hard}}^{\mathrm{timelike}}|/F_{\mathrm{hard}}^{\mathrm{spacelike}}$ is $1/\sqrt{1 + \pi^2/L^2}$, i.e., the timelike term is suppressed compared to the spacelike one.

Since the structure of the evolution corrections for the pion form factors is essentially identical to that of the $\gamma^*\gamma \to \pi^0$ form factor, to continue the evolution logarithms, we can use the approach outlined in Section III. The first step is to write the solution of the evolution equation [48,41,40] as an expansion over Gegenbauer polynomials $C_n^{3/2}(2x-1)$, the eigenfunctions of the LO kernel $V^{(0)}(x,y;\alpha_s)$:

$$\phi(x, Q^2) = \phi^{\mathrm{as}}(x) \left[ 1 + \sum_{k=1}^{\infty} a_{2k}^\pi(Q^2) C_{2k}^{3/2}(2x-1) \right] , \tag{49}$$

where $\phi^{\mathrm{as}}(x) = 6x(1-x)$ is the asymptotic distribution amplitude of the pion. The Gegenbauer moments $a_{2k}^\pi(Q^2)$ have a simple evolution pattern

$$a_{2k}^\pi(Q^2) = a_{2k}^\pi(\mu_0^2) \exp\left[-\gamma_{2k} \ln\left(Q^2/\mu_0^2\right)\right] \tag{50}$$

(we treat $\alpha_s$ as a constant here), with $\gamma_{2k}$ being the corresponding anomalous dimensions and $a_{2k}(\mu_0^2)$ the Gegenbauer moments of the initial distribution amplitude

$$\phi_0(x) = \phi^{\text{as}}(x)\left[1 + \sum_{k=1}^{\infty} a_{2k}^{\pi}(\mu_0^2) C_{2k}^{\frac{3}{2}}(2x-1)\right] . \tag{51}$$

This representation is very convenient to perform the analytic continuation to the timelike region of $Q^2$. Indeed, changing $Q^2 \to -q^2$, one obtains the natural shift $\ln(Q^2) \to \ln(q^2) + i\pi$, so that

$$a_{2k}^{\pi}(Q^2) \to a_{2k}^{\pi}(-q^2) = a_{2k}^{\pi}(|q^2|)e^{-i\delta_{2k}} , \tag{52}$$

where

$$\delta_{2k} \equiv \pi\gamma_{2k} . \tag{53}$$

From (52) it is obvious that the only interesting and potentially enhancing effect is due to the phases $\delta_{2k}$, since they can destroy some fine tuning of the coefficients $a_{2k}^{\pi}(\mu_0^2)$ and produce a positive interference. But in order to realize this possibility, one should start with a situation when there are negative coefficients, say, $a_2^{\pi}(\mu_0^2) < 0$, while the corresponding phase is close to $\pi$, e.g., $\delta_2 \approx \pi$. Such a situation is hard to imagine. Even the (unrealistic) CZ distribution amplitude has $a_2^{\pi}(\mu_0^2 = (0.5 \text{ GeV})^2) = 2/3$, while other models are closer to the asymptotic distribution amplitude, though all models provide $a_2 > 0$. Furthermore, the phase $\delta_2$ is $25/18\alpha_s$, so one needs a prohibitively large value $\alpha_s \sim 2.5$ for the coupling constant. In Fig. 3 we plot the ratio TL/SL for the pion form factor in the CZ model, taking the frozen value $\alpha_s(Q^2) = 0.3$. As one can see, the absolute value of $F_\pi^{\text{HARD}}(q^2)$ in the timelike region is reduced.

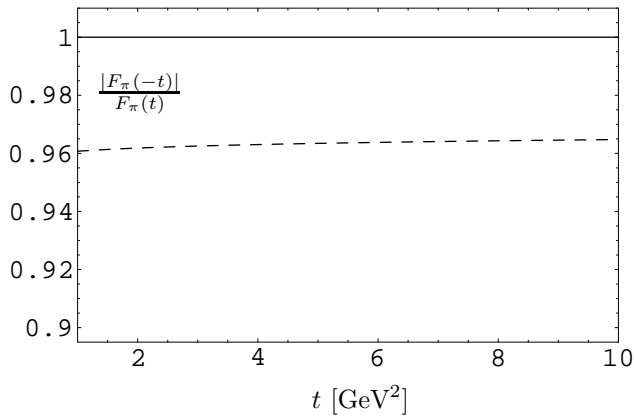

FIG. 3. The ratio of the timelike form factor $|F_\pi^{\text{HARD}}(q^2 = -t)|$ over the spacelike one $F_\pi^{\text{HARD}}(Q^2 = t)$ in the case of the CZ model (dashed line) for the pion distribution amplitude with $\alpha_s(Q^2)$ fixed to 0.3. The solid line represents the result for the asymptotic distribution amplitude.

To conclude, the perturbative contribution to the pion form factor, $F_\pi^{\text{HARD}}(Q^2)$, with a realistic distribution amplitude (which is close to the asymptotic one, see [63–66]) produces no sizable effects in analytically continuing to timelike $q^2$ values. The only potential effect is due to the substitution $\bar{\alpha}_s(Q^2) \to \tilde{\alpha}_s(q^2)$ which results in a 10%-reduction of the form factor.

## VI. EVOLUTION PHASES OF THE NUCLEON DISTRIBUTION AMPLITUDES AND THE NUCLEON FORM FACTOR

The nucleon form factor in the leading $\alpha_s$ order can be cast in the form [4]

$$Q^4 G_M^N(Q^2) = \frac{1}{54}\left[4\pi\bar{\alpha}_s(Q^2)\right]^2 |f_N|^2 \int_0^1 [dx] \int_0^1 [dy] \left[2\sum_{i=1}^{7} e_i T_i(x_j, y_j) + \sum_{i=8}^{14} e_i T_i(x_j, y_j)\right] , \tag{54}$$

where the amplitudes $T_i(x_j, y_j) = \phi^{(N)}(\{x\}, Q^2) T_H^i(\{x\}, \{y\}) \phi^{(N)}(\{y\}, Q^2)$ represent convolutions of $T_H^i$ with the appropriate distribution amplitudes $\phi^{(N)}(\{x\}, Q^2)$ evaluated term by term for each contributing diagram (marked by the index "$i$"). The nucleon distribution amplitude can be represented as an expansion over symmetrized combinations $\tilde{\Phi}(\{x\})$ of Appell polynomials (for more details, we refer to [6,67])

$$\phi^{(N)}(\{x\}, Q^2) = \phi^{\text{as}}(\{x\}) \sum_{n=0}^{\infty} B_n(Q^2) \tilde{\Phi}_n(\{x\}) , \qquad (55)$$

with

$$B_n(Q^2) = B_n(\mu_0^2) \exp\left[-\tilde{\gamma}_n \ln\left(Q^2/\mu_0^2\right)\right] , \qquad (56)$$

and eigenfunctions

$$\tilde{\Phi}_k(x_i) = \sum_{m,n=0}^{m+n=M} c_{mn}^k \mathcal{F}_{mn}(5, 2, 2; x_1, x_3) , \qquad (57)$$

where $\mathcal{F}_{mn}(5, 2, 2; x_1, x_3)$ are the Appell polynomials [‡]. Here

$$\tilde{\gamma}_n(M) = \frac{\alpha_s}{4\pi} \left(\frac{3}{2} C_F + 2\eta_n(M) C_B\right) , \qquad (58)$$

are the associated anomalous dimensions of trilinear quark operators with the quantum numbers of the nucleon containing external derivatives [67]. In Eq. (55) $\phi^{\text{as}}(\{x\}) = 120\, x_1 x_2 x_3$ denotes the asymptotic distribution amplitude of the nucleon and $B_n(\mu_0^2)$ are expansion coefficients for some initial distribution amplitude

$$\phi_0^{(N)}(\{x\}) = \phi^{\text{as}}(\{x\}) \sum_{n=0}^{\infty} B_n(\mu_0^2) \tilde{\Phi}_n(\{x\}) . \qquad (59)$$

Again, the representation given by Eq. (59) is very convenient to analyze the analytic continuation of the (hard part of) the nucleon form factor into the timelike region of $Q^2$. Continuing $Q^2 \to -q^2$ one obtains in Eq. (56) the same shift as in Eq. (52) with $\delta_n \equiv \alpha_s(\beta_0/4)\gamma_n$. Specifying the particular values of the coefficients $B_n$ we can calculate the ratio of timelike to the spacelike form factors for several models known in the literature. In Fig. 4 we display the ratio of the timelike to the spacelike form factors of the nucleon for three different nucleon distribution amplitudes: Chernyak–Ogloblin–Zhitnitsky, heterotic and Gari–Stefanis [6,69]. From this figure, it is clear that there is no enhancement due to the analytic continuation except from a marginal factor of order 1.03 for the GS model.

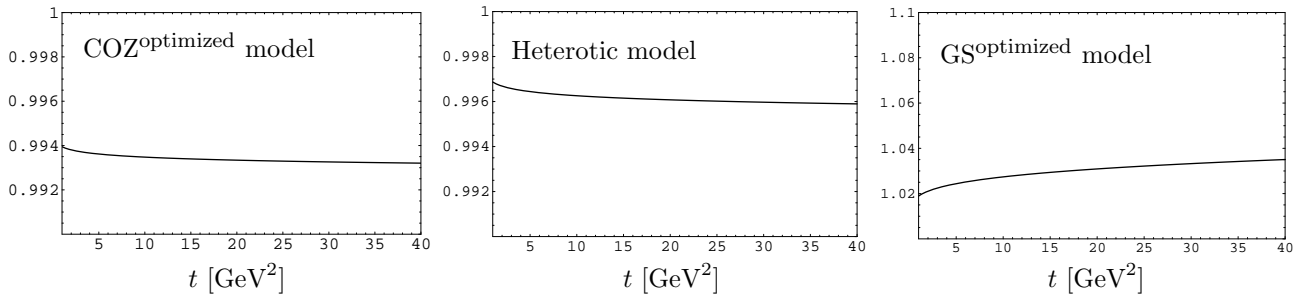

FIG. 4. Ratio of the timelike (with $t = q^2$) to the spacelike (with $t = Q^2$) hard form factors of the nucleon using three different nucleon distribution amplitudes (Chernyak–Ogloblin–Zhitnitsky, heterotic and Gari–Stefanis models).

---

[‡] One can also expand $\tilde{\Phi}_n(\{x\})$ over the polynomials proposed in Ref. [68]; the particular choice of the basis is not essential for our purposes.

# VII. SOFT TERMS FOR THE PION FORM FACTOR IN THE LOCAL DUALITY APPROACH

So far, we have discussed only the hard pQCD contributions to the hadronic form factors. But as argued in Refs. [19–25], the dominant contribution at intermediate values of the momentum transfers $Q^2 \leq 10$ GeV$^2$ is generated by the soft contribution which involves no hard gluon exchanges. As a model for the soft contribution, we assume the local duality (LD) approximation in which it is assumed that the pion form factor is dual to the free quark spectral density [21,70]

$$f_\pi^2 F_\pi(Q^2) = \frac{1}{\pi^2} \int_0^{s_0} \int_0^{s_0} \rho_0(s, s', Q^2) \, ds \, ds' \; . \tag{60}$$

The latter is given by [19–21]

$$\rho_0(s, s', t) = \frac{3}{4} \left[ t^2 \left(\frac{d}{dt}\right)^2 + \frac{1}{3} t^3 \left(\frac{d}{dt}\right)^3 \right] \frac{1}{\sqrt{(s + s' + t)^2 - 4ss'}} \; . \tag{61}$$

Here the duality interval $s_0$ corresponds to the effective threshold for the higher excited states and the "continuum" in the channels with the axial current quantum numbers.

In principle, the value of $s_0$ is fixed by the ratio of the nonperturbative power corrections to the (leading) perturbative term in the OPE for the correlator. In what follows, we use the value $s_0 \approx 0.7$ GeV$^2$ which has been extracted in the pioneering paper [71] from the QCD sum rule analysis of the correlator of two axial currents. The LD prescription for this correlator just implies the relation

$$s_0 = 4\pi^2 f_\pi^2 \; , \tag{62}$$

between $s_0$ and the pion decay constant $f_\pi$. This relation ensures that the Ward identity for the pion form factor

$$F_\pi(0) = 1 \tag{63}$$

is satisfied within the LD approach.

Performing the integral on the rhs of Eq.(60) we get the explicit analytic expression for the pion form factor

$$F_\pi^{\rm LD}(Q^2) = 1 - \frac{1 + 6s_0/Q^2}{(1 + 4s_0/Q^2)^{3/2}} \; , \tag{64}$$

originally obtained in [21,70]. Note that for $t \gtrsim 0.6$ GeV$^2$, expression (64) is in good agreement with existing data (see Fig. 5).

A simplified version of the LD model is based on using the "duality triangle" instead of the "duality square". In this approach [19,20,26], one uses the variables $S = s_1 + s_2$ and $s_1 - s_2$, introducing the reduced spectral density

$$\bar{\rho}_0(S, Q^2) \equiv \int_0^S \rho_0(S - s', s', Q^2) \, ds' \; . \tag{65}$$

The LD relation, eq. (60), is then substituted by its "triangle" version (TrLD)

$$F_\pi(Q^2) \simeq F_\pi^{\rm TrLD}(Q^2) = \frac{1}{\pi^2 f_\pi^2} \int_0^{S_0} \bar{\rho}_0(S, Q^2) \, dS \tag{66}$$

with $S_0 = \sqrt{2} s_0$. The latter condition means that the areas of the integration regions over $s$ and $s'$ in the two approaches are the same (see [20] and [26,21] for more details).

Using (61) and (65) we can easily calculate the spectral density $\bar{\rho}_0(S, Q^2)$

$$\bar{\rho}_0(S, Q^2) = \frac{S^2 \left(2S + 3Q^2\right)}{2 \left(2S + Q^2\right)^3} \tag{67}$$

producing

$$F_\pi^{\rm TrLD}(Q^2) = \frac{1}{\sqrt{2} \left(1 + Q^2/2S_0\right)^2} \; . \tag{68}$$

As one can see from the left part of Fig. 5 the difference between the two models in the region of interest ($Q^2 \gtrsim 1$ GeV$^2$) is very small.

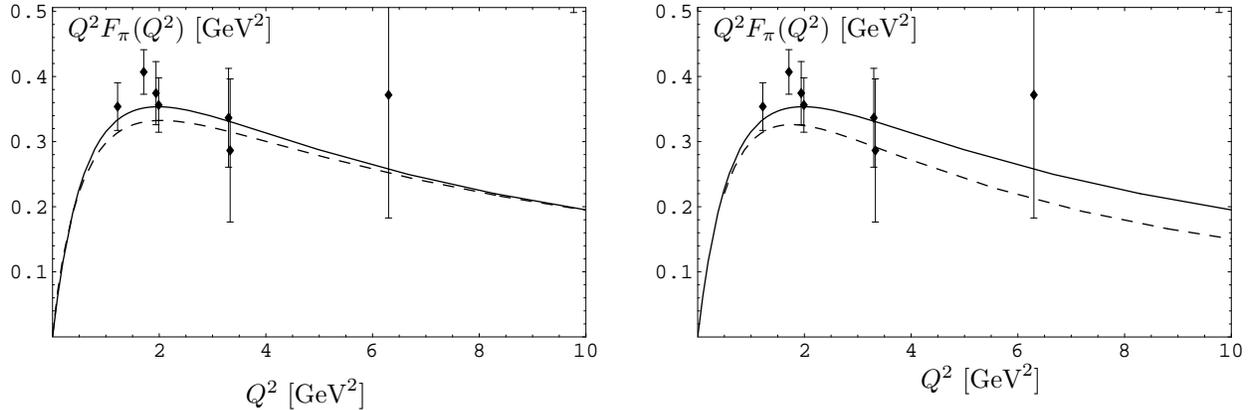

FIG. 5. Left part: Comparison of different LD models for the pion form factor. The solid line represents the triangle LD model and the dashed line the square LD model. Right part: Comparison with available experimental data from [72,73] of the pion form factor with (dashed line) and without (solid line) the Sudakov exponential in the 'triangle LD approach'.

### A. Sudakov effects due to the electromagnetic vertex in $F_\pi^{\rm TrLD}(Q^2)$

The crucial feature of the soft contribution is that it is accompanied by the Sudakov form factor. In other words, the double logarithms $\alpha_s \ln^2(Q^2)$ do not cancel in this case.

The one-loop radiative corrections to the spectral density $\bar\rho_0(S, t = Q^2)$ has been calculated by one of us (A.P.B.) [74] To analyze the Sudakov effects in the Feynman gauge, we need only the result for the gluon correction to the electromagnetic vertex (accompanied by the appropriate 1/2-insertions of self-energies into the quark lines, which gives an UV-finite result):

$$\frac{\Delta^{\rm EM\text{-}vertex}\bar\rho(S,t)}{(\alpha_s C_F/2\pi)\,\bar\rho_0(S,t)} = 2\left(\mathbf{Li}_2\left(\frac{S}{2S+t}\right) - \ln\left(1+\frac{t}{S}\right)\ln\left(2+\frac{t}{S}\right)\right)\left(1+\frac{t^2(6S+5t)}{S^2(2S+3t)}\right)$$
$$+ \left(\ln^2\left(2+\frac{t}{S}\right) - \mathbf{Li}_2(1)\right)\left(1+\frac{2t^2(6S+5t)}{S^2(2S+3t)}\right)\ln\left(2+\frac{t}{S}\right)\left(-\frac{1}{2}+\frac{2t(2S+5t)}{S(2S+3t)}\right) \quad (69)$$
$$+ 2\left(\mathbf{Li}_2\left(\frac{t}{2S+t}\right) - \ln\left(1+\frac{2S}{t}\right)\ln(2)\right)\frac{t^2(6S+5t)}{S^2(2S+3t)}\ln\left(1+\frac{2S}{t}\right)\frac{t^2(42S+55t)}{8S^2(2S+3t)} - \frac{5t}{4S}\,.$$

The leading asymptotics of this expression in the large $t$ regime is[§]

$$\frac{\Delta^{\rm EM\text{-}vertex}\bar\rho(S,t)}{\bar\rho_0(S,t)} \xrightarrow[t\to\infty]{} -\frac{\alpha_s C_F}{2\pi}\left[\ln\left(1+\frac{t}{2S}\right)\right]^2 + O\left(\ln\frac{t}{S}\right)\,. \quad (70)$$

So, we model the Sudakov corrections in the following way

$$\bar\rho_0^{\rm Sudakov}(S,Q^2) \simeq \bar\rho_0(S,Q^2)\exp\left[-\frac{\alpha_s C_F}{2\pi}\ln^2\left(1+\frac{Q^2}{2S}\right)\right]\,. \quad (71)$$

The modified spectral density is then used to model the soft term corrected by the Sudakov effects

$$F_\pi^{\rm TrLD-Sudakov}(Q^2) \equiv \frac{1}{\pi^2 f_\pi^2}\int_0^{S_0} \bar\rho_0^{\rm Sudakov}(S,Q^2)\,dS\,. \quad (72)$$

On the right part of Fig. 5 we show for comparison predictions for $Q^2 F_\pi^{\rm TrLD}(Q^2)$ and for $Q^2 F_\pi^{\rm TrLD\text{-}Sudakov}(Q^2)$. One can see from this figure that the Sudakov effects in the electromagnetic vertex reduce (as expected) the soft contribution in the spacelike region by 6–20%.

---

[§]A similar correction was obtained in the light-cone QCD sum rule approach [75].

## B. Model dependence of the soft term in timelike region

As we have seen in the previous subsection for spacelike values of the momentum transfer ($Q^2 > 0$) both LD models give rather close results for the pion form factor at $Q^2 \gtrsim 1$ GeV$^2$. But if we analytically continue these two models into the timelike region ($q^2 = -Q^2 > 0$), we obtain absolutely different results for both $\mathrm{Re}[F_\pi(q^2)]$ and $\mathrm{Im}[F_\pi(q^2)]$:

$$\mathrm{Re}[F_\pi^{\mathrm{LD}}(q^2)] = 1 - \theta(q^2 - 4s_0)\frac{1 - 6s_0/q^2}{(1 - 4s_0/q^2)^{3/2}} \ , \tag{73}$$

$$\mathrm{Im}[F_\pi^{\mathrm{LD}}(q^2)] = \theta(4s_0 - q^2)\frac{1 - 6s_0/q^2}{(4s_0/q^2 - 1)^{3/2}} \ , \tag{74}$$

$$\mathrm{Re}[F_\pi^{\mathrm{TrLD}}(q^2)] = \frac{1}{\sqrt{2}\,(1 - q^2/2S_0)^2} \ , \tag{75}$$

$$\mathrm{Im}[F_\pi^{\mathrm{TrLD}}(q^2)] = 0 \ . \tag{76}$$

We see that in the resonance region ($q^2 < 4$ GeV$^2$) the differences between these two models are rather large, and we can actually say nothing about the true behavior of $F_\pi(q^2)$ in this region. On the other hand, in the region $q^2 \gtrsim 6$ GeV$^2$ the differences between the two models are less than the experimental uncertainties, and hence we can use them, at least as a first approximation, to model $F_\pi(q^2)$. Furthermore, in the case of the 'triangle LD' we have an explicit analytic expression for the Sudakov effects which we can now continue into the timelike region

$$\mathrm{Re}[F_\pi^{\mathrm{TrLD-Sudakov}}(q^2)] = \frac{1}{\pi^2 f_\pi^2}\int_0^{S_0} \bar{\rho}_{0,\mathrm{Re}}^{\mathrm{Sudakov}}(S, q^2)\, dS \ ; \tag{77}$$

$$\mathrm{Im}[F_\pi^{\mathrm{TrLD-Sudakov}}(q^2)] = \frac{1}{\pi^2 f_\pi^2}\int_0^{S_0} \bar{\rho}_{0,\mathrm{Im}}^{\mathrm{Sudakov}}(S, q^2)\, dS \ , \tag{78}$$

with

$$\bar{\rho}_{0,\mathrm{Re}}^{\mathrm{Sudakov}}(S, q^2) = \tilde{\rho}_0(S, q^2)\cos\left(\tilde{\alpha}_s(q^2)C_F \ln\left[\frac{q^2}{2S} - 1\right]\right) \ ; \tag{79}$$

$$\bar{\rho}_{0,\mathrm{Im}}^{\mathrm{Sudakov}}(S, q^2) = \tilde{\rho}_0(S, q^2)\sin\left(-\tilde{\alpha}_s(q^2)C_F \ln\left[\frac{q^2}{2S} - 1\right]\right) \ ; \tag{80}$$

$$\tilde{\rho}_0(S, q^2) \equiv \bar{\rho}_0(S, -q^2)\exp\left[-\frac{\tilde{\alpha}_s(q^2)C_F}{2\pi}\left(\ln^2\left[\frac{q^2}{2S} - 1\right] - \pi^2\right)\right] \ .$$

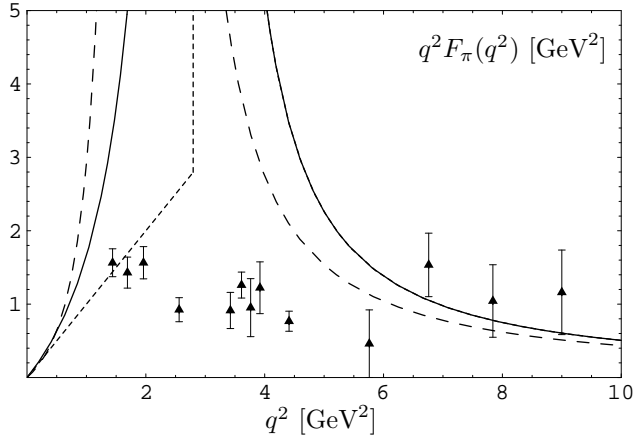

FIG. 6. Comparison of the analytic continuation to the timelike region $q^2 = -Q^2 > 0$ for two different LD models of the pion form factor. The solid line shows the result for $|q^2 F_\pi(q^2)|$, obtained from the 'square LD model', together with the real part of it (dotted line), whereas the prediction of the 'triangle LD model' for $q^2 F_\pi(q^2)$ is represented by the dashed line. The experimental data are taken from [16].

(Here $\tilde{\alpha}_s(q^2)$ is the lowest-order model expression (18) for $\alpha_s$ in the timelike regime.)

Using this model we obtain results, depicted on the lhs of Fig. 7. After adding the analytically continued expres-sion for the hard scattering (perturbative) part, including also transverse momentum effects (Sudakov+intrinsic effects) [76], we arrive at the result, shown on the rhs of Fig. 7.

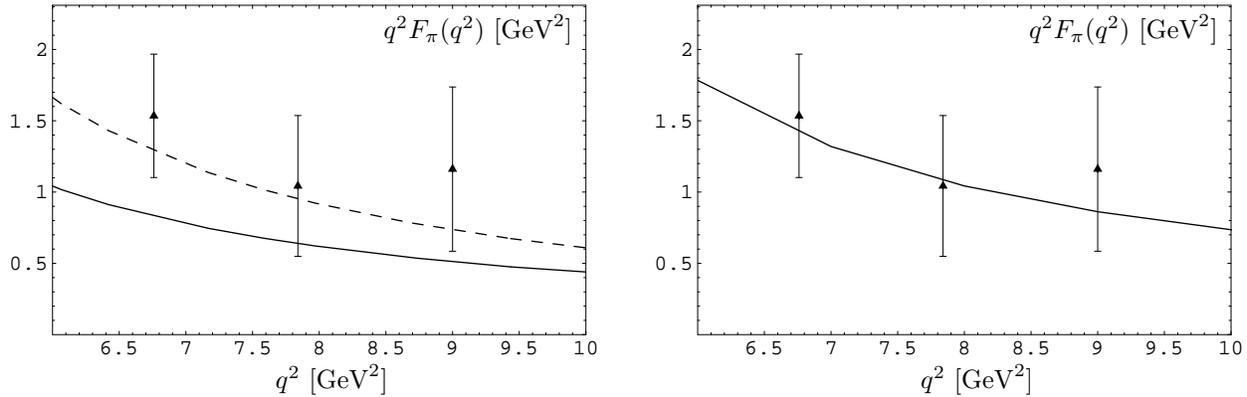

FIG. 7. The lhs shows the analytically continued expression for the pion form factor to the timelike region $q^2 = -Q^2 > 0$ with (dashed line) and without (solid line) the Sudakov exponential in the 'triangle LD approach'. The rhs shows the results for the total pion form factor in the timelike domain comprising the soft part within the 'triangle LD approach' and the hard one, calculated in [76]. Experimental data are taken from [16].

## VIII. CONCLUSIONS

In this paper, we investigated various aspects of the analytic continuation procedure from the spacelike to the timelike region of momentum transfers for several processes in QCD. We concentrated on studying several types of logarithmic contributions $\ln(Q^2)$ capable of producing $\pm i\pi$ in the timelike region. In the case of the ultraviolet logarithms, we reviewed the construction of the effective QCD coupling constant for the timelike region. The major result here is that the transition from a spacelike to the mirror timelike momentum only decreases the effective coupling constant. Studying the collinear logarithms, we established that in this case each eigenfunction $\phi_n(x)$ of the evolution equation acquires a phase factor $e^{i\delta_n}$. The phase vanishes for the asymptotic wave function, and there are no changes in this most realistic situation. But even in the case of the Chernyak-Zhitnitsky wave function, the interference effects are very small and, again, they decrease rather than increase the timelike contribution compared to the spacelike one. In the case of the pion electromagnetic form factor, we emphasized that the $\pi^2$ terms which may appear in the timelike region on the diagram by diagram level cancel in the total sum together with the double logarithms which generated them. Thus, we found no sources for the $K$-factor-type enhancements in the hard gluon exchange perturbative QCD contributions to the hadronic form factors. However, the situation completely changes if one considers the soft contribution. We investigated the simplest case of the pion electromagnetic form factor. To this end, we incorporated the local duality model suggested by the QCD sum rule studies performed earlier in the lowest (zero) order in $\alpha_s$. We included the $\alpha_s$ correction which, as expected, contains the Sudakov double logarithms. In the timelike region the latter produce non-canceling positive $\pi^2$ terms which result in a $K$-factor-type enhancement. Our results for the soft contribution are in good agreement with existing experimental data on the pion electromagnetic form factor both in the spacelike and the timelike regions. We regard this agreement as another indication that soft contributions dominate the form factors at currently accessible momentum transfers.

## ACKNOWLEDGMENTS

This work was supported in part by the US Department of Energy under contract DE-AC05-84ER40150; Russian Foundation for Fundamental Research (grant N 00-02-1669), Heisenberg–Landau Program and by the COSY Forschungsproject Jülich/Goeke. Two of us (A.B. and A.R.) are highly indebted to Prof. Klaus Goeke for the warm hospitality in Institut für Theoretische Physik II, Ruhr-Universität Bochum, where the major part of this work was done.